\documentclass[twocolumn,nofootinbib]{revtex4-1}

\pdfoutput=1

\usepackage{slashed}

\usepackage[normalem]{ulem}

\usepackage{color}

\usepackage{epsfig}
\usepackage{epstopdf}
\usepackage{epsf}
\input{epsf.sty}
\usepackage{graphicx,amsmath,amssymb}
\usepackage{epsfig}
\allowdisplaybreaks

\def\ei{\end{itemize}}
\def\be{\begin{equation}}
\def\ee{\end{equation}}
\newcommand{\bea}{\begin{eqnarray}}
\newcommand{\eea}{\end{eqnarray}}

\usepackage{bbm,bm,amsmath,amssymb}

\usepackage{hyperref}

\newcommand{\cF}{\mathcal{F}}
\newcommand{\cG}{\mathcal{G}}

\newcommand{\cH}{\mathcal{H}}
\newcommand{\cL}{\mathcal{L}}

\newcommand{\cN}{\mathcal{N}}

\renewcommand{\d}{\textrm{d}}

\renewcommand{\a}{\alpha}
\renewcommand{\l}{\lambda}
\renewcommand{\b}{\beta}

\def\E{{$E_{7(7)}$}}

\newcommand{\rf}[1]{(\ref{#1})}
\usepackage{float}

\begin{document}

\title{Is d=4 Maximal Supergravity Special?}
\author{Renata Kallosh}
\email{kallosh@stanford.edu}
\affiliation{Stanford Institute for Theoretical Physics and Department of Physics, Stanford University, Stanford,
CA 94305, USA}

 \begin{abstract}
We study  candidate counterterms (CT)'s  in maximal     supergravities in  diverse integer dimensions  $d \geq 4 $.
 We find that UV divergences  in  these theories occur at the number of loops $L$ below  certain critical value $L_{cr}$. 
  At $L\geq  L_{cr}$ the  CT's have nonlinear local supersymmetry and local $\cH$ symmetry, but the ones below $L_{cr}$ break these symmetries.  Therefore deforming the theories by adding the CT's  to the action to cancel UV divergences would be inconsistent with BRST.   Such  divergences with $L<  L_{cr}$ were found in  maximal supergravities  for all integer $d > 4$, but not for  $\cN=5,6,8$  in $d = 4$ so far, which renders the case $d = 4$ special.   \end{abstract}

\maketitle

 %\tableofcontents{}

%\newpage

\parskip 6pt

%%%%%%%%%%%%%%%%%%%%%%%%%%%%%%%%%%%%%%%%%%%%%%%%%%%%%%%%%%%%%%%%%%%%%%
%%%%%%%%%%%%%%%%%%%%%%%%%%%%%%%%%%%%%%%%%%%%%%%%%%%%%%%%%%%%%%%%%%%%%%
%%%%%%%%%%%%%%%%%%%%%%%%%%%%%%%%%%%%%%%%%%%%%%%%%%%%%%%%%%%%%%%%%%%%%%
%%%%%%%%%%%%%%%%%%%%%%%%%%%%%%%%%%%%%%%%%%%%%%%%%%%%%%%%%%%%%%%%%%%%%%
%%%%%%%%%%%%%%%%%%%%%%%%%%%%%%%%%%%%%%%%%%%%%%%%%%%%%%%%%%%%%%%%%%%%%%
\section{Introduction}

We will  study  here   loop computations  available in maximal supergravities in diverse dimensions. The physical scalars in these  theories are coordinates of the coset space ${\cG\over \cH}$. We will  discuss  the meaning of loop  computations with regard to the symmetries of  these supergravities: local nonlinear supersymmetry, local $\cH$-symmetry and global duality symmetry $\cG$. 

The main question we would like to address here is a possibility of a consistent deformation of these theories: can we  cancel UV divergences by adding to the theory UV divergent counterterms? For example, one can add the $R^3$ terms to pure gravity theory to absorb its 2-loop UV divergence. Can we do something similar in maximal supergravities?

In Sec. \ref{sec:II} we describe maximal  ${\cG\over \cH} $ supergravities in diverse dimensions, perform the dimensional analysis of UV divergences and compare this with the available computational data.

In \cite{Kallosh:2023asd,Kallosh:2023thj} we discussed a simple {\it dimensional analysis  of nonlinear local supersymmetry} (NLS), which is  basically a  dimensional analysis in superspace. It can be  applied in   perturbative supergravities at loop order $L<L_{cr}$. The full  analysis of a nonlinear local supersymmetry in combination with  duality symmetry  goes beyond  dimensional analysis and can be applied to loop order $L\geq L_{cr}$. In $d=4$ it was studied in \cite{Kallosh:2023asd}; here we give a short description of a full nonlinear local supersymmetry and duality in Sec. \ref{sec:GZ}. 

We show in this paper that that in all maximal supergravities in integer dimensions $d>4$ there are known UV infinities at 
loop order $L<L_{cr}$. Therefore here we will need to use only a dimensional analysis of  NLS for studies of maximal supergravities in diverse dimensions. In Sec. \ref{sec:IV} we summarize our findings.

In Appendix \ref{app:A} we also discuss the case of $d=5$ half-maximal supergravity since it is one of the  3 ``enhanced UV cancellation" cases presented in a recent review paper  \cite{Bern:2023zkg}. In particular, we explain an interesting result in \cite{Bern:2014lha} that at 4-loop order the 4-graviton amplitude is  UV finite, but there are some other 4-point amplitudes which are UV infinite.

In Appendix \ref{app:B} we show the difference between linear supersymmetry in amplitudes and   nonlinear supersymmetry  in supergravity Lagrangians, using the relevant superfields.

In expectation that eventually the computation of the UV properties in $\cN=5$ supergravity at $d=4$ at the   loop order $L=5$  will be performed, this analysis might be helpful to evaluate its potential meaning. What exactly would it mean if the UV divergence is found (or not) in this case from the point of view of the symmetries of the theory?

\section{\boldmath Maximal  ${\cG\over \cH} $ supergravities in diverse dimensions}\label{sec:II}
\subsection{\boldmath ${\cG\over \cH}$ coset space structure}
Duality symmetries in supergravities were discovered by Cremmer-Julia (CJ) and Gaillard-Zumino (GZ) in \cite{Cremmer:1979up,Gaillard:1981rj}. The corresponding \E\, duality symmetry $\cG$  is believed  to protect d=4 maximal supergravity from UV divergences. In diverse dimensions 
 duality symmetries of supergravities $\cG$ as well as local $\cH$-symmetries were studied in  \cite {Andrianopoli:1996ve,deWit:1997sz,Tanii:1998px}. 
 
\begin{figure}[h!]
\begin{center}
\includegraphics[scale=0.65]{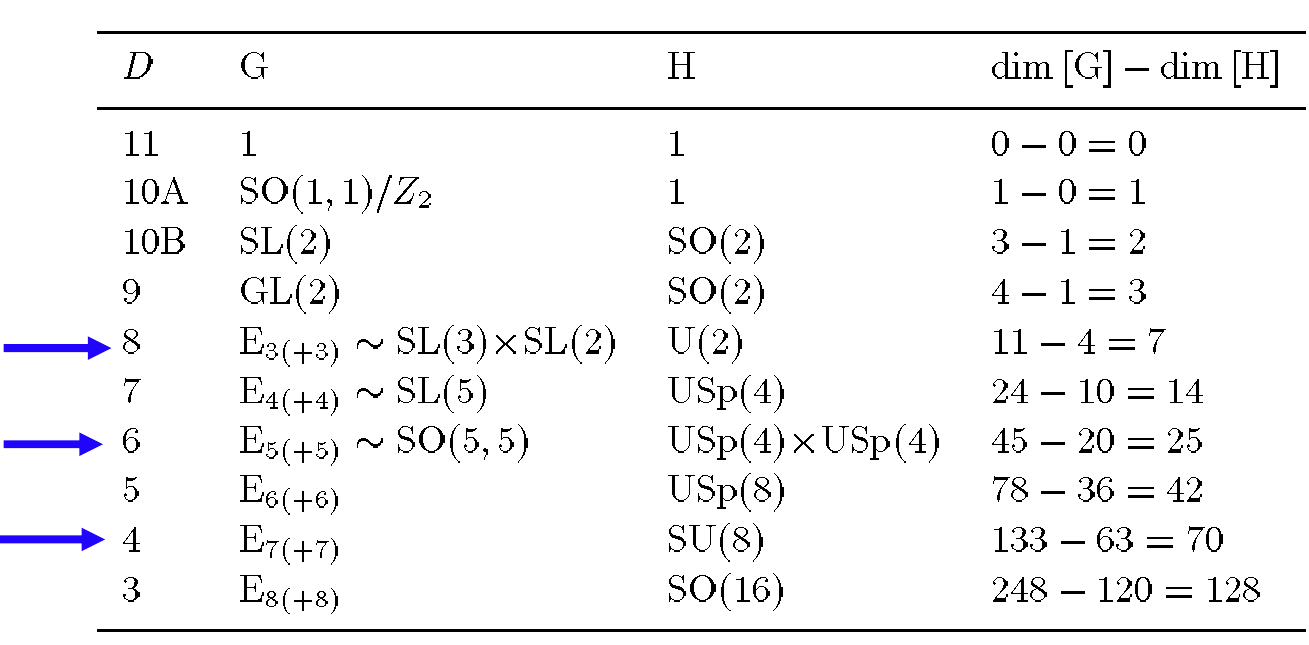}
\end{center}
\caption{\footnotesize Homogeneous scalar manifolds ${\cG\over \cH}$ for maximal supergravities in each integer dimensions from 11 to 3. The difference of dimensions of $\cG$ and of $\cH$ defines the number of scalars in each of these supergravities.
This is a Table from \cite{deWit:1997sz}. We have added blue arrows in d=4,6,8 where GZ type duality symmetry might control UV divergences.}
\label{GH}
\end{figure}

 In Fig. \ref{GH} here we present the table with maximal supergravities and their coset spaces from \cite{deWit:1997sz}.  We indicate dimensions $4,6,8$ maximal supergravities which have electro-magnetic dualities of the GZ type.
 
We will perform an analysis of UV divergences in supergravities in diverse dimensions. We will  compare the information we can extract from their global duality symmetry $\cG$, local supersymmetry  and local $\cH$-symmetry with
the data from actual computations of  loop diagrams \cite{Bern:1998ug,Bern:2007hh,Bern:2008pv,Bern:2009kd,Bern:2012uf,Bern:2018jmv}.
In $d=11$ both $\cG$ and of $\cH$ symmetries are trivial and there are no scalars. In $d=10$\,IIA there is no local $\cH$ symmetry and also duality flips it into $d=10$\,IIB theory, which does not have an action. Therefore our analysis will be valid 
for maximal supergravities in integer dimensions
\be
4\leq d \leq 9
\ee
The main focus here will be on dimensional analysis of NLS in $d\geq 5$, the details   in $d=4$ are in \cite{Kallosh:2023asd}.
When we study supergravity actions in diverse  dimensions and the symmetries of their actions and equations of motion,  fractional dimensions make no sense
 since explicit  actions and their symmetries  are available only in integer dimensions \cite{Salam:1989ihk}. 
 
 It was 
 suggested in  \cite{Herrmann:2018dja,Edison:2019ovj} that  d=4 maximal supergravity may have special properties at asymptotically large momenta, which may delay the UV infinities. Our analysis of symmetries of maximal supergravities in diverse integer dimensions  \cite{Salam:1989ihk} and comparison with their UV divergences known from  \cite{Bern:1998ug,Bern:2007hh,Bern:2008pv,Bern:2009kd,Bern:2012uf,Bern:2018jmv} also points out that  $d=4$ may be  special.

It is  stressed in  \cite{Edison:2019ovj} that the cancelations at infinity uncovered in their paper ``do not seem to be a consequence of gauge invariance or supersymmetry''.  But one should keep in mind that supersymmetry which one can see manifestly in superamplitudes corresponds to a linearized supersymmetry approximation of the full nonlinear supersymmetry of supergravity.  

In amplitudes there is a   super-Poincar\'e symmetry based on Lorentz symmetry of the space-time, not on general covariance,  see Appendix \ref{app:B} for details.

\subsection{Dimensional analysis of UV divergences}
First we remind the dimension of the  gravitational coupling.   The action is dimensionless
\be
S= {1\over \kappa^2} \int d^dx R = M_{Pl}^2 \int d^dx R \ .
\ee 
Therefore 
\be
{\rm dim} \Big [{1\over \kappa^2} \Big]= {\rm dim}  [M_{Pl}^2] = d-2  \ .
\ee
For integer dimensions $d\geq 5$ we define  
 the critical loop number as the one below which CT's have broken supersymmetry and  local $\cH$-symmetry.  The critical loop number is where CT's  preserve local $\cH$-symmetry and supersymmetry for the first time. 
 
We generalize   to  $d\geq 5$ the structure of the superinvariants \cite{Kallosh:1980fi,Howe:1980th}  based on the superspace  in $d=4$ \cite{Brink:1979nt, Howe:1981gz}.  We conclude that the superinvariants might have unbroken supersymmetry and 
  and local $\cH$-symmetry  if  1) we integrate over the whole superspace volume 2) the minimal dimension superfield $\cH$-invariant Lagrangian depends on 4 spinor superfields.  Therefore the Lagrangian has a minimum  dimension  2   and the superinvariant for the case of $L=L_{cr}$ is
\be
S_{cr} = \kappa^{2 \, (L_{cr}-1)} \int d^{4\cN} \d^d x \, det\,  E\, \cL(x, \theta) \ .
\ee 
In $d=4$ this is the expression proposed in \cite{Kallosh:1980fi,Howe:1980th} as a  geometric candidate CT which starts at loop order $L=\cN$. In more general dimensions we find that 
\be
{\rm dim} [\kappa^2] \, (L_{cr}-1) + {\rm dim}\,  [\rm full \, \, superspace] +{\rm dim}\,[\cL]=0 \ .
\ee
Now we have
\be
-(d-2) (L_{cr}-1)+ 2\cN -d+2 =0 \ ,
\ee
and we conclude that  $L_{cr}$ is (when integer)
\be
L_{cr}={2\cN\over d-2} \ ,
\label{crInt}\ee
or  (when ${2\cN\over d-2}$ is not an integer)
\be
L_{cr}=\Big [{2\cN\over d-2}\Big ] +1\, .  
\label{cr}\ee
Here $\Big [{2\cN\over d-2}\Big ]$ is the  integer part of a  number, defined as  the part of the number that appears before the decimal.

In $d=4$, where
 one can expects new  computational data for $\cN=5,6,8$,  the critical value is\footnote{There was a claim in \cite{Bossard:2011tq}  that  in $d= 4$ that there are nonlinear supersymmetric CT's  in the harmonic superspace at $L=\cN-1$.     We  study the case $d=4$ separately in
\cite{Kallosh:2023asd}.}
\be
 L_{cr}^{d=4} =\cN \ ,
\label{d4}\ee
in agreement with \cite{Kallosh:1980fi,Howe:1980th}.
For maximal supergravities with $\cN=8$ we find according to \rf{crInt} and \rf{cr} 
\be
    L_{cr}^{max}={16\over d-2}\, ,  \quad {\rm or} \quad L_{cr}^{max}=\Big [{16\over d-2}\Big ] +1\, .  
\label{crM}\ee
In particular,
\bea
 L_{cr}^{d=4, \, \cN=8} =8\, , \quad   L_{cr}^{d=5, \, \cN=8} =6 \, , 
\quad L_{cr}^{d=6,  \, \cN=8} =4\, , \cr
\cr
 L_{cr}^{d=7,  \, \cN=8} =4\, , \quad L_{cr}^{d=8,  \, \cN=8} =3\, , \quad L_{cr}^{d=9,  \, \cN=8} =3\, .  \nonumber 
\eea
In terms of curvatures the first expression which has a nonlinear supersymmetric embedding is
\bea 
 &d=4, \, L_{cr}=8 :  \quad \kappa^{14} \int d^4\, x \, D^{10} R^4\, ,   \quad &{\rm dim} [\kappa^2]  =  -2 \cr
&d=5,  \, L_{cr}=6 :  \quad \kappa^{10} \int d^5\, x \, D^{12} R^4\, ,   \quad &{\rm dim} [\kappa^2]  = -3 \cr
& d=6,  \, L_{cr}=4 :  \quad \kappa^{6} \int d^6\, x \, D^{10} R^4\, ,   \quad &{\rm dim} [\kappa^2]  =  -4 \cr
& d=7,  \, L_{cr}=4 :  \quad \kappa^{6} \int d^7\, x \, D^{14} R^4\, ,   \quad &{\rm dim} [\kappa^2]  = - 5 \cr
& d=8,  \, L_{cr}=3 :  \quad \kappa^{4} \int d^8\, x \, D^{12} R^4\, ,   \quad &{\rm dim} [\kappa^2]  =  -6\cr
& d=9,  \, L_{cr}=3 :  \quad \kappa^{4} \int d^9\, x \, D^{15} R^4\, ,   \quad &{\rm dim} [\kappa^2]  =  -7  \nonumber
\eea
The nature of this critical loop order can be also explained by reminding that to have  $L < L_{cr}$ we have to restrict to   submanifolds of the superspace, i. e. to integrate over the number of $\theta$'s which is less than $4\cN$. In addition to integration over less that $4\cN$  $\theta$'s one typically needs to take
 the Lagrangians of dimension zero, depending on non-geometric superfields with the first component being a scalar field. 

The most important feature of all  $L<L_{cr}$ candidate CT's is that they have only linear supersymmetry and are available only in the unitary gauge where the local $\cH$-symmetry is already gauge-fixed. They do not exist in case of nonlinear supersymmetry and before gauge-fixing. For these we have to look at $L\geq L_{cr}$.

\subsection{Data on UV divergences from computations}
In \cite{Bern:1998ug,Bern:2007hh,Bern:2008pv,Bern:2009kd,Bern:2012uf,Bern:2018jmv} we find  that
the first divergence in maximal supergravities  have been found  at $5\leq d\leq9$
\bea
&&d = 9, \qquad L_{UV} = 2 <  L_{cr}=3 \cr
&&d = 8, \qquad L_{UV} = 1 <  L_{cr}=3 \cr
&&d = 7, \qquad L_{UV}= 2 < L_{cr}=4\cr
&&d=6,   \qquad L_{UV}=3 <L_{cr}=4\cr
&&d=5,   \qquad L_{UV}>4\, , \, \, \, \, L_{cr}=6
\eea
 One  could   read the results in \cite{Bern:2018jmv} as a statement that  since $5>{24\over 5}= d_c$, the $d=5$ maximal supergravity is UV divergent at $L=5$. If this is the case we add that 
\be
\hskip  0.5 cm d=5, \qquad L_{UV}=5 < L_{cr}=6 \ .
\ee
It appears that in all cases where we have the information about UV divergences  in $d\geq 5$ from loop computations, the UV divergences were discovered at the loop order below what we have defined as $L_{cr}$. This means that the relevant candidate CT's in all these maximal supergravities  break full nonlinear supersymmetry and local $\cH$-symmetry.

It means that they cannot be viewed as viable terms which would deform the action to absorb UV divergences: they do not have local symmetries of the classical action.

The cases with even $d=6,8$ are particularly interesting to consider in more detail since they have specific ``electro-magnetic'' GZ-type duality symmetry of the kind we encounter in $d=4$. So we may learn more about the protective power of duality from these examples. Like the case $\cN=5, d=4, L=\cN-1=4$ where computations are available, suggests some insight into   $\cN=8, d=4, L=\cN-1=7$ where the computations are not available.

\section{GZ duality symmetries in various dimensions in maximal supergravities}\label{sec:GZ}
\subsection{Electro-magnetic duality}
Electro-magnetic symmetries of maximal supergravity in d=4 were discovered by Cremmer-Julia (CJ) and Gaillard-Zumino (GZ) in \cite{Cremmer:1979up,Gaillard:1981rj}. The corresponding \E\, symmetry is believed  to protect d=4 maximal supergravity from UV divergences. In diverse dimensions 
 electro-magnetic type duality symmetries  were studied in  \cite {Andrianopoli:1996ve,Tanii:1998px,deWit:1997sz}.
 
It was explained in  \cite {Andrianopoli:1996ve}  that GZ type duality symmetry is possible  in $d> 4$  only in $d=6$ and $d=8$ maximal supergravities. The relevant  $p$-extended object is dual to the magnetic one with  ${d\over 2}= p+2$.  In odd dimensions d=5,7,9 there are only electric charges, no magnectic ones, so duality symmetry does not involve the Noether-Gaillard-Zumino current conservation. In d=10IIA supergravity duality relates it to d=10IIB, which is also not of the GZ type duality of a single supergravity theory.

The basic feature of GZ duality, following  \cite {Andrianopoli:1996ve,Tanii:1998px,deWit:1997sz},   is that in even dimensions $d=2n$ there is an $(n-1)$-th rank antisymmetric tensor field $B^I_{\mu_1\dots \mu_{n-1}}$ $(I=1, \dots M$) and there is a field strength  and its dual, both of rank $n$ ($n=2,3,4$ for $d=4,6,8$ respectively)
\bea
F^I_{\mu_1\dots \mu_{n}}= n \, \partial_{[\mu_1} B^I_{\mu_2\dots \mu_{n}]}\, , \nonumber \\
\tilde F^{ I \mu_1\dots \mu_{n}}= {1\over n!}  \, \epsilon ^{ \mu_1\dots \mu_{n} \nu_1\dots \nu_{n}}    F^I_{\nu_1\dots \nu_{n}} \ .
\eea 
Duality groups $\cG$ in maximal supergravities in $d=4,6,8$ are 
\be
 \cG= E_{11-d(11-d)} \, : \quad E_{7(7)}, \quad \quad  E_{5(5)} , \quad \quad E_{3(3)}  \ ,
\ee
respectively, see also Fig. \ref{GH}. The $\cH$ local symmetry groups  in these theories are
\be
 \cH  \, : \qquad SU(8), \quad USp(4) \times USp(4), \quad U(2) \ .
\ee
The equations of motion for $B^I_{\mu_1\dots \mu_{n-1}}$ and the Bianchi identities are
\be
\partial_{\mu_1} ( e \tilde G_I^{\mu_1\dots \mu_{n}})=0 \, , \qquad \partial_{\mu_1} ( e \tilde F_I^{\mu_1\dots \mu_{n}})=0 \ ,
\ee
where
\be
\tilde G_I ^{\mu_1\dots \mu_{n}} = {n!\over e} {\partial \cL \over \partial F^I_{\mu_1\dots \mu_{n}}}\, , \quad G= \eta \tilde {\tilde G}\, , \quad \eta = (-1) ^{{d\over 2} +1} \ .
\label{G}\ee
Vector equations of motion and Bianchi identities are invariant under the transformations mixing $F$ and $G$
\begin{equation}
\delta_{\cG} \left ( \begin{array}{c}   F\cr   G\cr   \end{array}\right
) \, =\, \left ( \begin{array}{cc} A & B \cr C & D \cr  \end{array} \right )
\left (  \begin{array}{c}  F\cr  G\cr  \end{array}\right ).
\label{dualrot}
\end{equation}
The  constant matrix ${\cal S} =\left (\begin{array} {cc}A & B \cr C & D
\cr  \end{array} \right ) \, \in \, Sp(2M,\mathbb{R} )$ or its subgroup for $d= 4k$ or to $SO(M, M)$ for $d=4k+2$.

With regard to  UV divergences, it is important to stress that classical duality symmetry for $F$ involves ${\partial \cL^{cl} \over \partial F^I_{\mu_1\dots \mu_{n}}}
$. Therefore, when CT's are added to the action, duality symmetry has to be deformed as it will involve ${\partial (\cL^{cl} +\l \cL^{CT})\over \partial F}
$ and duality transformations  become $\l$-dependent
\be
\delta_{\cG} F=  \delta F^{cl}+ \l B \hat G= \delta F^{cl} + \lambda B \eta  {n!\over e}  {\partial  \cL^{CT} \over \partial \tilde F}  \ .
\label{cGdef}\ee
Deformation of the bosonic action of  $d=4$ supergravities due to this deformation of duality symmetry was studied in \cite{Kallosh:2018mlw}.

More symmetries in supergravities are affected by the deformation of the action. Specifically, fermions which are neutral on duality symmetry $\cG$  transform only under local $\cH$ symmetry. Under supersymmetry they  transform via a  field strength which is a combination of scalars and $F$ and $G$ which is $\cG$ invariant. In d=4 this combination is known as a graviphoton 
\be
\cF_{\mu \nu}^{ij} (F, G) \, , \qquad \bar \cF_{\mu \nu \, ij} (F, G) \ ,
\ee
which is neutral under \E\, duality  and transforms under local $SU(8)$.

In $d=6$ it is a combination of a 3-form $F$ and a 3-form $G$, defined in eq. \rf{G}. In $d=8$ it is a combination of a 4-form $F$ and a 4-form $G$,  where $G$ defined in eq. \rf{G}. We will use a notation 
$
\cF (F, G)
$ for these $\cG$-invariant, $\cH$-covariant combination of vectors field strength's $F$ and their duals $G$ defined in eq. \rf{G}.

When the CT is added to the classical action, the local supersymmetry of fermions has to be deformed, so that after supersymmetry transformation the fermion is still duality invariant.
 It acquires a $\l$-dependent term, same as the duality transformation $\cG$ acting on antisymmetric tensor field strength's $F$ in eq. \rf{cGdef}.
\be
\delta_S ( \chi )= \delta_S (\chi^{cl}) + \l  {\partial  \cL^{CT} \over \partial  \cF} \epsilon \ .
\label{fdef}\ee
Here $\epsilon$ is a local supersymmetry parameter. Details on this and a  specific example of the deformation of supersymmetry of the fermions in $d=4$ is presented in \cite{Kallosh:2023asd}. Thus, $d=4, 6, 8$ with $\cN=8$ have certain GZ duality in  common.

In $d=8$ there is $L=1$ UV divergence. The corresponding CT breaks local supersymmetry and local $U(2)$ symmetry. We do not even need to study the effect of GZ duality in this case. If we would have the first UV infinity at $L_{cr}=3$ where we, at least in principle, anticipate the existence of the geometric superinvariant with on shell $E_{3(3)}$ and local $U(2)$ symmetry, we would have an example of the case where one can assume the deformation via $\cL^{CT}$ is possible, and study the consequences of deformation of duality and supersymmetry, as in eqs. \rf{cGdef}, \rf{fdef}. 

But since at $L=1$ there is no eligible CT, we can only conclude that in this case the computation has revealed that UV infinity is breaking local symmetries of the theory. It makes the corresponding perturbative QFT of deformed maximal supergravity in $d=8$ BRST inconsistent. 

In $d=6$ there is $L=3$ UV divergence \cite{Bern:2008pv} below $L_{cr}=4$. It means that the  corresponding CT breaks local supersymmetry and local $USp(4)\times USp(4)\sim SO(5)\times SO(5)$ symmetry. As in $d=8$ case with its 1-loop UV divergence, there is no need to study the effect of GZ duality in this case. If we would have the first UV infinity at $L_{cr}=4$ where we, at least in principle, anticipate the existence of the geometric superinvariant with on shell $E_{5(5)}$ and local $USp(4)\times USp(4)$ symmetry, we would have an example of the case where one can assume the deformation via $\cL^{CT}$ is possible, and study the consequences of deformation of duality and supersymmetry, as in eqs. \rf{cGdef}, \rf{fdef}. 

But since at $L=3$ there is no eligible CT, we can only conclude that in this case the computation has revealed that UV infinity is breaking local symmetries of the theory. It makes the corresponding perturbative QFT of maximal supergravity in $d=6$ BRST inconsistent. 

\subsection{\boldmath $d=6$ $L=3$ UV divergence and its absence at $d=4$ $L=3$}
Maximal supergravity in $d=4$ $L=3$ is UV finite \cite{Bern:2007hh} whereas maximal supergravity in $d=6$ $L=3$ is UV infinite  \cite{Bern:2008pv}. If, as it is believed, $d=4$ case is protected by \E\, symmetry, why is $d=6$  not protected by $E_{5(5)}$ symmetry?

One of the reasons to study UV divergences in diverse dimension was to understand the case of $d=6$ maximal supergravity and the role of GZ duality \cite{Cremmer:1979up,Gaillard:1981rj}. It was suggested\footnote{ H. Nicolai, private communication}   that maximal supergravity 
in $d=6$ may be viewed as an analog of a maximal supergravity in $d=4$: instead of  vectors in $d=4$ in $d=6$ there are five 2-form fields $A_{mn}$ such that $F_{mnp}= \partial_{[m} A_{np]}$ and
${\delta L\over \delta F_{mnp}}$  together form a 10-plet of $SO(5,5) \sim E_{5(5)}$. Therefore one would have expected that  $E_{5(5)}$ Noether-Gaillard-Zumino current  conservation would protect the theory from UV divergences. However, since there is a UV divergence in $d=6$ at $L=3$ one can view it as  a counterexample to the case of  \E\, symmetry at $d=4$. 

Now we have learned  that to sort out the role of \E\, symmetry at $d=4$ versus $E_{5(5)}$ in $d=6$ one has to look more carefully at the relevant candidate CT's.  In $d=4$ the geometric superinvariants start at $L_{cr}=8$ whereas in $d=6$ they start at $L_{cr}=4$.  The UV divergence at $L=3, d=6$ can be related to a linearized ${1\over 8}$ BPS superinvariant in harmonic space \cite{Bossard:2009sy}, symbolically
\be
\kappa^{4} \int d^6 x D^6 R^4  \rightarrow \kappa^4 \int d^6 x \, du \, (d^{28} \theta)_1 (W^4)_1 \ .
\ee
This superinvariant is available only in the unitary gauge where local  $USp(4)\times USp(4)\sim SO(5)\times SO(5)$ symmetry is gauge-fixed, and only at the linearized supersymmetry level. This superinvariant when added to a classical action, breaks both local  $USp(4)\times USp(4)$ symmetry as well as a  local nonlinear supersymmetry. 

Whether there is also some kind of 1-loop  anomaly  in maximal $d=6$ supergravity remains to be seen. 

 At $L=3, d=4$ the 1/2  of the superspace superinvariant is available \cite{Kallosh:1980fi}. In harmonic space it is  ${1\over 2} $ BPS superinvariant \cite{Bossard:2009sy}
\be
\kappa^{4} \int d^4 x  R^4  \rightarrow \kappa^4 \int d^4 x \, du \, (d^{16} \theta)_{1234} (W^4)_{1234} \ .
\ee 
It is also available only in the unitary gauge where local $SU(8)$ symmetry is gauge-fixed, and only at the linearized supersymmetry level. This superinvariant if added to a classical action, would break local  $SU(8)$ symmetry and nonlinear local supersymmetry. But since there is no UV divergence, there is no need for such a deformation.

Therefore the UV divergence below $L_{cr}$ in both cases indicates that the corresponding CT lacks nonlinear supersymmetry and local $SU(8)$ symmetry in $d=4$ and local $USp(4) \times USp(4)\sim SO(5)\times SO(5)$ symmetry in $d=6$.
The UV divergence in $d=6$ at $L=3$ is explained as the one which is not supported by a valid CT with nonlinear supersymmetry.  In $d=4$ $\cN=8$ case this would be the case of UV divergences at $L=3, \dots , 7$. So far we have not seen it. 

A related important case is $d=4$, $\cN=5$, 
$L=4$, which is UV finite. A crucial information where we reach the critical level loop order will come from  $d=4$, $\cN=5$, 
$L=5$.

\section{Discussion}\label{sec:IV}

In this paper we have shown that all cases of  maximal supergravities\footnote{The case with half-maximal supergravity in diverse dimensions needs to be studied separately. Here we have only added  in Appendix \ref{app:A} a case with enhanced cancellation in $d=5$ half-maximal supergravity at 2 loops. It is   one of the three enhanced cases discussed in the review article \cite{Bern:2023zkg}.  We have  noticed a similarity between half-maximal supergravities in $d=4$ and $d=5$, both are UV divergent at one loop above the UV finite enhanced case, and UV divergences include anomalous terms  breaking nonlinear local supersymmetry.} in integer dimensions $4< d<10$ have UV divergences at the loop order which is below $L_{cr}$, where for the first time the eligible candidate CT's are available. We define {\it eligible CT's here as the ones which can be added to the action and deform it while preserving the gauge symmetries of the classical action, namely nonlinear local supersymmetry and local $\cH$-symmetry}.

Our results follow from dimensional analysis in superspace for dimensions $d=5,6,7,8,9$. Comparing this dimensional analysis with loop computations in \cite{Bern:1998ug,Bern:2007hh,Bern:2008pv,Bern:2009kd,Bern:2012uf,Bern:2018jmv} we conclude that all maximal supergravities in these dimensions are UV divergent at the loop level where the relevant superinvariants are not eligible candidate CT's: they
break local symmetries of the classical theories. This implies that the deformed action of $d=5,6,7,8,9$ maximal supergravities of the form $S^{def}= S^{cl} + \l S^{CT}$, capable of absorbing UV divergences, are BRST inconsistent. This means that 
 one cannot cancel UV divergences in $d=5,6,7,8,9$ maximal supergravities without  breaking their gauge theory consistency.
 
  It appears therefore that it is preferable to 
study  perturbative $d=4$, $\cN\geq 5$ supergravities, which are  not  known to have anomalies \cite{Freedman:2017zgq} and/or UV divergences breaking nonlinear local supersymmetry.

\

\noindent{\bf {Acknowledgments:}} I am grateful to Z. Bern, J. J. Carrasco, H. Elvang, D. Freedman, A. Linde, H.~Nicolai, R. Roiban  and  Y. Yamada  for stimulating discussions. 
 This work is supported by SITP and by the US National Science Foundation grant PHY-2014215.

\appendix
\section{Enhanced cancellation in $d=5$ half-maximal supergravity at 2 loops}\label{app:A}
\subsection{d=4 half-maximal supergravity}
In $d=4$ the absence of the 3-loop UV divergence  is explained using dimensional analysis of NLS in \cite{Kallosh:2023asd}: $L=3< L_{cr}=\cN=4$. At $L=4$ however, there are UV divergences of 3 types \cite{Bern:2013uka}.
The corresponding  $L=\cN=4 $   superinvariants, candidate for the UV divergence which involve a 4-graviton amplitude are known and expected for 40 years
\cite{Kallosh:1980fi}
 \be
 {\rm CT}_1^{L=4} \to  \kappa^{6} \int d^4 x\, d^{8} \theta \, d^{8} \bar \theta\,  \det E \bar \chi_{\dot \alpha i}   \bar \chi^{\dot \alpha}_{ j} \chi_{\alpha}^{ i}  \chi^{\alpha j} \,,
\label{I4}\ee
but the second and the third terms, ${\rm CT}_2^{L=4} $ and ${\rm CT}_3^{L=4} $ have  helicity structure of the $U(1)$ anomalous amplitudes \cite{Carrasco:2013ypa}
\be
 {\rm CT}_2^{L=4} \to  \kappa^{6} \int d^4 x\, d^{8} \theta \, \bar C_{\dot \alpha \dot \beta \dot \gamma \dot \delta} W \partial^{\alpha \dot \alpha} \partial^{\beta \dot \beta} W \partial_{\alpha}^{ \dot \gamma} \partial_{\beta}^{ \dot \delta} W + hc \,,
\label{I43}\ee
\be
 {\rm CT}_3^{L=4} \to  \kappa^{6} \int d^4 x\, d^{8} \theta \, W^2 \partial^6 W^2 + hc\,,
\label{I42}\ee

The corresponding  UV divergences/$U(1)$ anomalous superamplitudes \rf{I42}, \rf{I43}  do not include a 4-graviton amplitude. This is easy to see from the properties of the superfields involved in \rf{I42}, \rf{I43} as opposite to \rf{I4} where 4-graviton amplitude is present.

 These 3  UV divergences  in  the amplitude language have the following  
  helicity structure 
\bea
&&{\rm CT}_1 \, :  \,  {\cal O}^{- -+ +} = 4s^2 t {\langle 12\rangle ^2\over [12] \langle 23\rangle \langle 34\rangle \langle 41\rangle} \cr
\cr
&&{\rm CT}_2 \, :  \,  {\cal O}^{- + + +}= -12 s^2 t^2 {[ 24]^2\over [ 12] \langle 23\rangle \langle 34\rangle [ 41]} \cr
\cr
&&{\rm CT}_3 \, :  \,  {\cal O}^{+ + + +} = 3st (s+t){ [12]  [34]\over \langle 12\rangle \langle 34\rangle}
\label{helicity}\eea
Each enters in the UV divergence \cite{Bern:2013uka} with a non-vanishing factor since each of the CT's has a linearized supersymmetry and fits by dimension to $L=4$.
\subsection{d=5 half-maximal supergravity}

It was found in computations in \cite{Bern:2012gh} that  $d=5$ half-maximal supergravity at 2 loops is UV finite, despite the CT available in
\cite{Bossard:2013rza}. Let us apply the formula in \rf{cr} to this case 
\be
L_{cr}=\Big [{2\cN\over d-2}\Big ] +1 = 3\, , \quad \rightarrow \quad   L_{cr}=\Big [{8 \over 3}\Big ] +1 = 3 \ .
\label{cr1}\ee
Thus, the explanation of the enhanced cancellation in this case is that the UV divergence/CT would  break a simple NLS, if it would show up in these 2-loop computations, which is one loop below the critical one.
\be
L=2<L_{cr}=3 \ .
\ee
We can compare it with the expression in eq. (7.28) in \cite{Bossard:2013rza}. It can be used as a 2-loop candidate CT
\be
\kappa^2 \int d^5 x \, d^{16} E \Phi = {9\over 32} \int d\mu_{(4,1)} \epsilon^{\a\b \gamma \delta} \chi _\a \chi_\b \chi_{\gamma} \chi_{\delta} \ .
\label{Bos}\ee
It is either an integral over the full superspace with the zero dimension superfield Lagrangian, or an integral over (harmonic) subspace of the superspace. Both  of these are valid only at the linear level. Either way, the CT breaks nonlinear supersymmetry. Therefore the enhanced cancellation observed in \cite{Bern:2012gh} is explained by an unbroken simple nonlinear local supersymmetry.

At the 3-loop order half-maximal supergravity was investigated in \cite{Bern:2014lha}. The result was given in the following form: there is  the 3-loop UV divergence in $d_c=14/3$. One can try to interpret this result as follows\footnote{I am grateful to J. J. Carrasco for the help with this interpretation.}. Since $d=5>14/3$ we assume that the property of the UV divergence at  $d_c=14/3$ can be valid also for $d=5$ case. 

Since in our analysis we are using the actual Lagrangian\footnote{E. Cremmer, Supergravities in 5 dimensions, in: ``Superspace and Supergravity'', proceedings of the Workshop held in July 1980 in Cambridge, England, eds. S.W. Hawking and M. Rocek, Cambridge University Press, p.267.} of the half-maximal supergravity and the relevant superspace described in \cite{Bossard:2013rza} we can compare the answers in \cite{Bern:2014lha} with the structure of the candidate UV divergences for $L=3$. This procedure is the same we will  explained above  in $d=4$ case.

We look at eq. (5.4) in \cite{Bern:2014lha} trying to interpret it as a structure of UV divergence analogous to the one in $d=4$. We find the following 3 structures
\bea\label{norm}
&&{\cal P}_{- -+ +}=0\\
\cr \label{anom1}
&& {\cal P}_{- + + +}= -{1\over 8} s^2 t^2 {[24]^2\over [12]\langle 23\rangle  \langle34\rangle [41]}\\
\cr
&& {\cal P}_{+ + + +} = {5\over 4} stu {[12] [34]\over \langle 12\rangle  \langle34\rangle }
\label{helicity5}\eea
If our interpretation that the $d_c=14/3$ result in  \cite{Bern:2014lha} can be valid for $d=5$ is correct, we should be able to explain the UV divergences in eqs. \rf{norm}, \rf{anom1}, \rf{helicity5} using the symmetries of the half-maximal supergravity in $d=5$. This is, indeed, possible.

The zero in the case \rf{norm} follows from the fact that at 3-loop order we need the following superinvariant, which would contain a 4-graviton amplitude
\be
\kappa^{2(L-1)} \int d^5 x \, d^{12} \theta \epsilon^{\a\b \gamma \delta} \chi _\a \chi_\b  D^{3(L-2)}\chi_{\gamma} \chi_{\delta} \ .
\ee
 At $L=2$ it will give us \rf{Bos} in the form of a subspace of the full superspace linearized invariant
$
\kappa^{2} \int d^5 x \, d^{12} \theta \epsilon^{\a\b \gamma \delta} \chi _\a \chi_\b  \chi_{\gamma} \chi_{\delta} ] .
$. For $L=3$ we are looking at 
\be
\kappa^{4} \int d^5 x \, d^{12} \theta \epsilon^{\a\b \gamma \delta} \chi _\a \chi_\b  D^3 \chi_{\gamma} \chi_{\delta}  ,
\label{3loop}\ee
so that $-2(3) -5 +6 +2+3=0$. 

To keep the 4-point graviton amplitude we need to use only space-time derivatives in $D^3$, but this will not give us a scalar Lagrangian. The reason is that spin 1/2 superfield has gravity in its 3d component with $\theta^3$ so that all 12 $\theta$'s will have all 4-gravitons. 

Any other superfield, a fermionic derivative of  spin 1/2 superfield, will have other fields in the 4-point amplitude, but will not have all 4 gravitons.

This explains the zero in \rf{norm} nicely. Moreover, it supports our interpretation of the result in $d_c=14/3$  in  \cite{Bern:2014lha} as being valid also in $d=5$ where we actually can make an analysis and predictions based on the Lagrangian in $d=5$ and on properties of superspace in \cite{Bossard:2013rza}.

Two other structures of 3-loop UV divergences in eqs. \rf{anom1}, \rf{helicity5} are associated with linearized superinvariants of the form \rf{3loop} where $D^3$ involves some of the fermionic derivatives. Alternatively, one can use a linearized superinvariant for the 3-loop UV divergence of the form of the integral over all superspace
\be
\kappa^4 \int d^5 x \, d^{16} E \, \cL (x, \theta) \ ,  \label{ano}\ee
where the superfield Lagrangian has dimension 3: it can depend on various linearized superfields of the theory  of different dimensions, starting from zero. 

In this form again one can see that 4-graviton amplitude will be absent. One can find  explicit linearized superinvariants which would fit the expressions for amplitudes in  eqs. \rf{anom1}, \rf{helicity5}.

In conclusion, the available computation of the 3-loop UV divergences in $d_c=14/3$ result in  \cite{Bern:2014lha} has been interpreted here as describing the UV divergences in $d=5$. The CT, corresponding to \rf{norm} is shown to be absent using the properties of $d=5$ supergravity, supporting  the result in  \cite{Bern:2014lha}. 

Two other UV divergences break NLS since the relevant CT's we presented in eqs. \rf{3loop} and \rf{ano} do not have an embedding into a full nonlinear superinvariant.

Thus, after a nice 2-loop enhanced cancellation of UV divergences in half-maximal $d=5$, $L=2$ supergravity, we see that at $L=3$ the UV divergences are present and moreover, they break local nonlinear supersymmetry. 

This is in a complete analogy with analogous  enhanced cancellation case in $d=4, L=3$  where in  $d=4, L=4$ there are UV divergences and they break  local nonlinear supersymmetry.

\section{Linear supersymmetry in amplitudes,   nonlinear  in supergravity Lagrangians}\label{app:B}

The best way to see the difference between these two is to look at eqs. (8.21)-(8.25) in  \cite{Cremmer:1979up} for nonlinear local supersymmetry transformation of the Lagrangian in eqs.  (8.16) where the notations are explained in eqs. (8.17)-(8.19).

The complexity of supergravity  actions  and its local supersymmetries prevented them from being useful for loop computations. But since the double copy method of computations nicely described in  \cite{Bern:2023zkg} is expected to give the result one could have obtained from direct supergravity computations, at this stage it is useful to clearly distinguish linearized supersymmetries manifest in double copy method and nonlinear local supersymmetry of supergravity Lagrangians.

A simple way to see this non-linearity is to start with the supersymmetry variation of the spin 1/2 fermion $\lambda$ in eq. (8.25) in   \cite{Cremmer:1979up} where it flips into a graviphoton $\cF$. Now we use a covariant derivative acting on (8.23) and learn that graviphoton $\cF$ is flipped into a field strength of gravitino. Now using (8.24) one can see that the local supersymmetry of gravitino field strength flips it into a  full nonlinear Riemann-Christoffel curvature, shown in red below
\be
\delta_s D_{[\nu} \psi_{\mu] A}=[D_\nu, D_\mu]_A{}^B \epsilon _B \label{curv}\ee
We take into account that the index $A$ represent a spinorial one $\alpha$ as well as an internal one $i$
\bea
&& \hskip -30pt \delta_s D_{[\nu} \psi_{\mu] \alpha i}   =  [D_\nu, D_\mu]_{\alpha i}{}^{\beta j} \epsilon _{\beta j} +\dots \nonumber \\
&&\hskip 20pt = {\color{red}{R_{\mu\nu \lambda \delta}}}\, 
(\sigma^{\lambda \delta}) _\alpha{}^\beta \epsilon _{\beta i} + (R_{\mu\nu})_i{}^j \epsilon_{\alpha j } +\dots
\label{curv}\eea
Now compare this with the superamplitude constructions based on a superfield (super-wave function) in eq. (12.25) of \cite{Elvang:2013cua}
\be
\Phi_i = h^+ + \eta_{iA} \psi^A +\dots + \prod _{k=1}^8 \eta_{ik} h^- \ .
\label{HY}\ee
The physical state of the graviton is flipped by linear supersymmetry into a physical state of gravitino etc all the way to a single graviton of opposite helicity. But the superfield in \rf{HY} is linear in each field. This is in agreement with the properties of super-Poincare generators - momentum $P^{\dot \a \b}$,  and the supercharges $Q^A$  and $\tilde Q_A$ so that 
\be
\{Q^{\dot \a i} , \tilde Q_{\b j}\} = P^{\dot \a \b} \delta^i{}_j
\label{Q} \ee
and momenta $P^{\dot \a \b} \rightarrow \partial ^{\dot \a \b}$ commute
\be
\{P^{\dot \a \b} , P^{\dot \gamma \delta }\} = 0 \qquad \Rightarrow \qquad \{\partial_\mu , \partial _\nu \}=0 \ .
\label{Lor}\ee
It is explained in \cite{Elvang:2013cua} that in $\cN=8$ supergravity the definition of super-Poincar\'e generators acting on superamplitudes, is different from the ones in $\cN=4$ super-Yang-Mills theory, given in their eqs (5.1) and (4.35), only by the fact that internal indices run from 1 to 8 instead of 1 to 4. This means that the space-time has  Lorentz symmetry but not general covariance, as it is also clear from \rf{Lor}. Nonlinear supersymmetry means that instead of \rf{Q}, \rf{Lor} we have
a commutator of 2 nonlinear supersymmetries  producing not a derivative (a  Poincar\'e translation) but a general coordinate covariant derivative. The anti-commutator of the fermionic derivatives generates a covariant space-time derivative
\be
\{D^{\dot \a i} , D_{\b j }\} = D_{\dot \a \b}  \delta^i{}_j+\dots 
\ee
The commutator of these covariant derivatives gives a space-time curvature.
With $ D_\mu = D_{\dot \a \b}\sigma^{\dot \a \b}_\mu $ 
\be
\{D_\mu , D_\nu \}  X_\l = R_{\mu\nu\l \delta}  X^\delta \ ,
\ee
replacing eq. \rf{Lor} of linearized supersymmetry in superampltudes.

We have also explained in   \cite{Kallosh:2023asd} around  eq. (1.12)   the meaning of the nonlinear supersymmetry of the Lagrangian, translated into a superfield language. It means   that the third 
  component of the  superfield $\lambda_{\a \, ijk} (x, \theta) $   is a Weyl spinor $C_{\a \b \gamma\delta}(x)$. 
  \be
D_\a^i D_\b^j  D_\gamma^k \,  \l_{\delta \, ijk}(x, \theta) |_{\theta=0} = C_{\a \b  \gamma\delta} (x) \ .
\ee
Weyl spinor is algebraically related to a nonlinear Riemann-Christoffel tensor in eq. \rf{curv} and both involve an infinite series in the graviton $h_{\mu\nu}$ when expanded near the flat space where $g_{\mu\nu}= \eta_{\mu\nu} +h_{\mu\nu}$.

\bibliography{refs}

\providecommand{\href}[2]{#2}\begingroup\raggedright\begin{thebibliography}{10}

\bibitem{Kallosh:2023asd}
R.~Kallosh and Y.~Yamada, ``{Deformation of d=4, N\ensuremath{>} 4
  Supergravities Breaks Nonlinear Local Supersymmetry}'',
  \href{http://arxiv.org/abs/2304.10514}{{\tt arXiv:2304.10514 [hep-th]}}.

\bibitem{Kallosh:2023thj}
R.~Kallosh, ``{Explaining enhanced UV divergence cancellations}'',
  \href{http://arxiv.org/abs/2304.13211}{{\tt arXiv:2304.13211 [hep-th]}}.

\bibitem{Bern:2023zkg}
Z.~Bern, J.~J.~M. Carrasco, M.~Chiodaroli, H.~Johansson, and R.~Roiban,
  ``{Supergravity amplitudes, the double copy and ultraviolet behavior}'',
  \href{http://arxiv.org/abs/2304.07392}{{\tt arXiv:2304.07392 [hep-th]}}.

\bibitem{Bern:2014lha}
Z.~Bern, S.~Davies, and T.~Dennen, ``{The Ultraviolet Critical Dimension of
  Half-Maximal Supergravity at Three Loops}'',
  \href{http://arxiv.org/abs/1412.2441}{{\tt arXiv:1412.2441 [hep-th]}}.

\bibitem{Cremmer:1979up}
E.~Cremmer and B.~Julia, ``{The SO(8) Supergravity}'',
\href{http://dx.doi.org/10.1016/0550-3213(79)90331-6}{{\em Nucl. Phys.} {\bf
  B159} (1979)  141--212}.
%%CITATION = NUPHA,B159,141;%%.

\bibitem{Gaillard:1981rj}
M.~K. Gaillard and B.~Zumino, ``{Duality Rotations for Interacting Fields}'',
\href{http://dx.doi.org/10.1016/0550-3213(81)90527-7}{{\em Nucl. Phys.} {\bf
  B193} (1981)  221--244}.
%%CITATION = NUPHA,B193,221;%%.

\bibitem{Andrianopoli:1996ve}
L.~Andrianopoli, R.~D'Auria, and S.~Ferrara, ``{U duality and central charges
  in various dimensions revisited}'',
  \href{http://dx.doi.org/10.1142/S0217751X98000196}{{\em Int. J. Mod. Phys.}
  {\bf A13} (1998)  431--490},
\href{http://arxiv.org/abs/hep-th/9612105}{{\tt arXiv:hep-th/9612105
  [hep-th]}}.
%%CITATION = HEP-TH/9612105;%%.

\bibitem{deWit:1997sz}
B.~de~Wit and J.~Louis, ``{Supersymmetry and dualities in various
  dimensions}'', {\em NATO Sci. Ser. C} {\bf 520} (1999)  33--101,
  \href{http://arxiv.org/abs/hep-th/9801132}{{\tt arXiv:hep-th/9801132}}.

\bibitem{Tanii:1998px}
Y.~Tanii, ``{Introduction to supergravities in diverse dimensions}'', in {\em
  {YITP Workshop on Supersymmetry}}.
\newblock 2, 1998.
\newblock \href{http://arxiv.org/abs/hep-th/9802138}{{\tt
  arXiv:hep-th/9802138}}.

\bibitem{Bern:1998ug}
Z.~Bern, L.~J. Dixon, D.~C. Dunbar, M.~Perelstein, and J.~S. Rozowsky, ``{On
  the relationship between Yang-Mills theory and gravity and its implication
  for ultraviolet divergences}'',
  \href{http://dx.doi.org/10.1016/S0550-3213(98)00420-9}{{\em Nucl. Phys. B}
  {\bf 530} (1998)  401--456}, \href{http://arxiv.org/abs/hep-th/9802162}{{\tt
  arXiv:hep-th/9802162}}.

\bibitem{Bern:2007hh}
Z.~Bern, J.~J. Carrasco, L.~J. Dixon, H.~Johansson, D.~A. Kosower, and
  R.~Roiban, ``{Three-Loop Superfiniteness of N=8 Supergravity}'',
  \href{http://dx.doi.org/10.1103/PhysRevLett.98.161303}{{\em Phys. Rev. Lett.}
  {\bf 98} (2007)  161303},
\href{http://arxiv.org/abs/hep-th/0702112}{{\tt arXiv:hep-th/0702112
  [hep-th]}}.
%%CITATION = HEP-TH/0702112;%%.

\bibitem{Bern:2008pv}
Z.~Bern, J.~J.~M. Carrasco, L.~J. Dixon, H.~Johansson, and R.~Roiban,
  ``{Manifest Ultraviolet Behavior for the Three-Loop Four-Point Amplitude of
  N=8 Supergravity}'', \href{http://dx.doi.org/10.1103/PhysRevD.78.105019}{{\em
  Phys. Rev. D} {\bf 78} (2008)  105019},
  \href{http://arxiv.org/abs/0808.4112}{{\tt arXiv:0808.4112 [hep-th]}}.

\bibitem{Bern:2009kd}
Z.~Bern, J.~J. Carrasco, L.~J. Dixon, H.~Johansson, and R.~Roiban, ``{The
  Ultraviolet Behavior of N=8 Supergravity at Four Loops}'',
  \href{http://dx.doi.org/10.1103/PhysRevLett.103.081301}{{\em Phys. Rev.
  Lett.} {\bf 103} (2009)  081301},
\href{http://arxiv.org/abs/0905.2326}{{\tt arXiv:0905.2326 [hep-th]}}.
%%CITATION = ARXIV:0905.2326;%%.

\bibitem{Bern:2012uf}
Z.~Bern, J.~J.~M. Carrasco, L.~J. Dixon, H.~Johansson, and R.~Roiban,
  ``{Simplifying Multiloop Integrands and Ultraviolet Divergences of Gauge
  Theory and Gravity Amplitudes}'',
  \href{http://dx.doi.org/10.1103/PhysRevD.85.105014}{{\em Phys. Rev. D} {\bf
  85} (2012)  105014}, \href{http://arxiv.org/abs/1201.5366}{{\tt
  arXiv:1201.5366 [hep-th]}}.

\bibitem{Bern:2018jmv}
Z.~Bern, J.~J. Carrasco, W.-M. Chen, A.~Edison, H.~Johansson,
  J.~Parra-Martinez, R.~Roiban, and M.~Zeng, ``{Ultraviolet Properties of
  $\mathcal N = 8$ Supergravity at Five Loops}'',
  \href{http://dx.doi.org/10.1103/PhysRevD.98.086021}{{\em Phys. Rev. D} {\bf
  98} (2018) no.~8, 086021}, \href{http://arxiv.org/abs/1804.09311}{{\tt
  arXiv:1804.09311 [hep-th]}}.

\bibitem{Salam:1989ihk}
A.~Salam and E.~Sezgin, eds., \href{http://dx.doi.org/10.1142/0277}{{\em
  {Supergravities in Diverse Dimensions}: {Commentary and Reprints (In 2
  Volumes)}}}.
\newblock World Scientific, Singapore, 1989.

\bibitem{Herrmann:2018dja}
E.~Herrmann and J.~Trnka, ``{UV cancellations in gravity loop integrands}'',
  \href{http://dx.doi.org/10.1007/JHEP02(2019)084}{{\em JHEP} {\bf 02} (2019)
  084}, \href{http://arxiv.org/abs/1808.10446}{{\tt arXiv:1808.10446
  [hep-th]}}.

\bibitem{Edison:2019ovj}
A.~Edison, E.~Herrmann, J.~Parra-Martinez, and J.~Trnka, ``{Gravity loop
  integrands from the ultraviolet}'',
  \href{http://dx.doi.org/10.21468/SciPostPhys.10.1.016}{{\em SciPost Phys.}
  {\bf 10} (2021) no.~1, 016}, \href{http://arxiv.org/abs/1909.02003}{{\tt
  arXiv:1909.02003 [hep-th]}}.

\bibitem{Kallosh:1980fi}
R.~E. Kallosh, ``{Counterterms in extended supergravities}'',
\href{http://dx.doi.org/10.1016/0370-2693(81)90964-3}{{\em Phys. Lett.} {\bf
  B99} (1981)  122--127}.
%%CITATION = PHLTA,B99,122;%%.

\bibitem{Howe:1980th}
P.~S. Howe and U.~Lindstrom, ``{Higher Order Invariants in Extended
  Supergravity}'',
\href{http://dx.doi.org/10.1016/0550-3213(81)90537-X}{{\em Nucl. Phys.} {\bf
  B181} (1981)  487--501}.
%%CITATION = NUPHA,B181,487;%%.

\bibitem{Brink:1979nt}
L.~Brink and P.~S. Howe, ``{The $N=8$ Supergravity in Superspace}'',
\href{http://dx.doi.org/10.1016/0370-2693(79)90464-7}{{\em Phys. Lett.} {\bf
  88B} (1979)  268--272}.
%%CITATION = PHLTA,88B,268;%%.

\bibitem{Howe:1981gz}
P.~S. Howe, ``{Supergravity in Superspace}'',
\href{http://dx.doi.org/10.1016/0550-3213(82)90349-2}{{\em Nucl. Phys.} {\bf
  B199} (1982)  309--364}.
%%CITATION = NUPHA,B199,309;%%.

\bibitem{Bossard:2011tq}
G.~Bossard, P.~S. Howe, K.~S. Stelle, and P.~Vanhove, ``{The vanishing volume
  of D=4 superspace}'',
  \href{http://dx.doi.org/10.1088/0264-9381/28/21/215005}{{\em Class. Quant.
  Grav.} {\bf 28} (2011)  215005},
\href{http://arxiv.org/abs/1105.6087}{{\tt arXiv:1105.6087 [hep-th]}}.
%%CITATION = ARXIV:1105.6087;%%.

\bibitem{Kallosh:2018mlw}
R.~Kallosh, H.~Nicolai, R.~Roiban, and Y.~Yamada, ``{On quantum compatibility
  of counterterm deformations and duality symmetries in $ \mathcal{N}\ge 5 $
  supergravities}'', \href{http://dx.doi.org/10.1007/JHEP08(2018)091}{{\em
  JHEP} {\bf 08} (2018)  091},
\href{http://arxiv.org/abs/1802.03665}{{\tt arXiv:1802.03665 [hep-th]}}.
%%CITATION = ARXIV:1802.03665;%%.

\bibitem{Bossard:2009sy}
G.~Bossard, P.~S. Howe, and K.~S. Stelle, ``{The Ultra-violet question in
  maximally supersymmetric field theories}'',
  \href{http://dx.doi.org/10.1007/s10714-009-0775-0}{{\em Gen. Rel. Grav.} {\bf
  41} (2009)  919--981}, \href{http://arxiv.org/abs/0901.4661}{{\tt
  arXiv:0901.4661 [hep-th]}}.

\bibitem{Freedman:2017zgq}
D.~Z. Freedman, R.~Kallosh, D.~Murli, A.~Van~Proeyen, and Y.~Yamada, ``{Absence
  of U(1) Anomalous Superamplitudes in $\mathcal{N}\geq 5$ Supergravities}'',
  \href{http://dx.doi.org/10.1007/JHEP05(2017)067}{{\em JHEP} {\bf 05} (2017)
  067},
\href{http://arxiv.org/abs/1703.03879}{{\tt arXiv:1703.03879 [hep-th]}}.
%%CITATION = ARXIV:1703.03879;%%.

\bibitem{Bern:2013uka}
Z.~Bern, S.~Davies, T.~Dennen, A.~V. Smirnov, and V.~A. Smirnov, ``{Ultraviolet
  Properties of N=4 Supergravity at Four Loops}'',
  \href{http://dx.doi.org/10.1103/PhysRevLett.111.231302}{{\em Phys. Rev.
  Lett.} {\bf 111} (2013) no.~23, 231302},
\href{http://arxiv.org/abs/1309.2498}{{\tt arXiv:1309.2498 [hep-th]}}.
%%CITATION = ARXIV:1309.2498;%%.

\bibitem{Carrasco:2013ypa}
J.~J.~M. Carrasco, R.~Kallosh, R.~Roiban, and A.~A. Tseytlin, ``{On the U(1)
  duality anomaly and the S-matrix of N=4 supergravity}'',
  \href{http://dx.doi.org/10.1007/JHEP07(2013)029}{{\em JHEP} {\bf 07} (2013)
  029},
\href{http://arxiv.org/abs/1303.6219}{{\tt arXiv:1303.6219 [hep-th]}}.
%%CITATION = ARXIV:1303.6219;%%.

\bibitem{Bern:2012gh}
Z.~Bern, S.~Davies, T.~Dennen, and Y.-t. Huang, ``{Ultraviolet Cancellations in
  Half-Maximal Supergravity as a Consequence of the Double-Copy Structure}'',
  \href{http://dx.doi.org/10.1103/PhysRevD.86.105014}{{\em Phys. Rev. D} {\bf
  86} (2012)  105014}, \href{http://arxiv.org/abs/1209.2472}{{\tt
  arXiv:1209.2472 [hep-th]}}.

\bibitem{Bossard:2013rza}
G.~Bossard, P.~S. Howe, and K.~S. Stelle, ``{Invariants and divergences in
  half-maximal supergravity theories}'',
  \href{http://dx.doi.org/10.1007/JHEP07(2013)117}{{\em JHEP} {\bf 07} (2013)
  117}, \href{http://arxiv.org/abs/1304.7753}{{\tt arXiv:1304.7753 [hep-th]}}.

\bibitem{Elvang:2013cua}
H.~Elvang and Y.-t. Huang, ``{Scattering Amplitudes}'',
  \href{http://arxiv.org/abs/1308.1697}{{\tt arXiv:1308.1697 [hep-th]}}.

\end{thebibliography}\endgroup
%\bibliography{refsN}
\bibliographystyle{utphys}
\end{document}